\documentstyle[12pt,a4]{article}

\begin{document}
\oddsidemargin=-.3cm
\evensidemargin=-1cm

\begin{titlepage}
\begin{center}
\vskip .5cm

\hfill CERN--TH/95--87\\
\hfill hep-th/yymmxxx\\

\vskip 4cm

{\large \bf Bianchi-type string cosmology}

\vskip .5cm

{\bf  Nikolaos A. Batakis}
\footnote{Permanent address: Department of Physics, University of
Ioannina,
GR--45100
 Ioannina, Greece}
\footnote{e--mail address: batakis@surya20.cern.ch}

{\em Theory Division, CERN\\
     CH--1211 Geneva 23, Switzerland}

\vskip 2cm

{\bf abstract }
\end{center}
\begin{quotation}\noindent
\end{quotation}
\noindent Bianchi-type string cosmology involves  generalizations
of the FRW backgrounds with three transitive spacelike Killing
symmetries, but without any {\em a priori} assumption of isotropy
in the 3D sections of homogeneity. With emphasis on those cases with
diagonal metrics and vanishing cosmological constant
which have not been previously examined in the
literature, the present findings allow an overview and the
classification of all Bianchi-type backgrounds. These string solutions
(at least to lowest order in $\alpha^\prime$) offer prototypes for
the study of spatial anisotropy and its impact on the dynamics of
the early universe.

\vskip 3cm
\noindent
CERN--TH/95--87 \\
April  1995\\
\end{titlepage}

\newpage
\section{Introduction}

Bianchi-type string cosmology involves 4D spatially homogeneous
spacetimes which satisfy at least the lowest-order
string beta-function eguations \cite{1}--\cite{7}. Disposing with the
assumption of spatial isotropy, these Bianchi-type string backgrounds
(BTSB) generalize (and contain as a special case) all
possible Friedmann-Robertson-Walker models
as well as those with asymptotic or less than $SO(3)$ isotropy.
As such, they provide the best models available
for the understanding of anisotropy and its impact on the
dynamics of the early universe,
well before the attainment of the presently observed state
of isotropy \cite{1},\cite{5}--\cite{7}. It is in this
vaguely-understood region where most of
the fundamental cosmological problems arise and also where string
cosmology has its best chance of being confronted with reality.

Bianchi-type cosmologies \cite{1},\cite{5}--\cite{7}
can be generally defined in terms of a
3-parameter group of isometries $G_3$.  All nine possible
group {\em types} are classified in terms of
the parameter $X=I,II,\ldots,IX$ and they
are further characterized as being of $G_3$-class $\cal A$ or $\cal B$
according to whether their adjoint representation is traceless or not
\cite{1}. The action of $G_3$ is simply transitive on its orbits so that
each orbit, equivalently identified with a 3D hypersurface of homogeneity
$\Sigma^3$, is spanned by three independent spacelike Killing vectors
$\xi_i$. The set of their duals $\{\sigma^i\}$,
invariant under the left action of $G_3$,
provides a natural (non-holonomic) basis for the formulation
of $G_3$-invariant statements. For example, the characterization
of a Bianchi-type metric as
{\em diagonal} presumes the employment of a
$\{\sigma^i\}$ basis because, for most types, such a
metric will not remain diagonal when expressed in terms of ordinary
(holonomic) coordinates.

As we will see, all BTSBs with diagonal metrics can be assembled in
three classes. To facilitate the discussion (and anticipate the
classification introduced later on) we denote these
classes as $X(d\uparrow)$, $X(d\rightarrow)$, $X(d\nearrow)$. The arrows
specify the orientation of the dual $H^\ast$ of the
totally antisymmetric field strength with respect to the (pictured
as `horizontal') $\Sigma^3$ sections. The $X(d\uparrow)$,
recently discussed in \cite{6}, contains as a subclass all possible
FRW backgrounds with vanishing cosmological constant $\Lambda$ and
generalizes all such previously known BTSBs \cite{3}--\cite{5}.
Also recenty discussed was the $X(d\rightarrow)$ class \cite{7}. It
follows that, with the investigation of the remaining $X(d\nearrow)$
case, the category $X(d)$ of all $\Lambda=0$ diagonal BTSBs can
be fully uncovered. Subsequently, an overview and a classification
of all possible BTSBs could be attained. These last
remarks also describe the motivation and objective of this paper.

In the following section we introduce notation
and certain preliminaries needed for
the presentation of our main results in section 3.
These are further discussed in section 4, which also contains
a classification (with brief reviews) on all possible diagonal
$\Lambda=0$ BTSBs and a summarizing Table.

\section{Preliminaries}

We consider 4D spacetimes with Bianchi-type metrics
of the form \cite{1},\cite{5}--\cite{7}
\begin{equation}
ds^2=-dt^2+a_1^2(t)(\sigma^1)^2+a_2^2(t)(\sigma^2)^2+
a_3^2(t)(\sigma^3)^2, \label{met}
\end{equation}
namely diagonal in the $\{dt,\sigma^i\}$ basis, as part of a
background solution which satisfies at least the lowest-order
string beta-function equations
for conformal invariance. The metric coefficients $a_i$ are functions
of the cosmic time $t$ only and, as mentioned, $\{\sigma^i\}$ is a
$G_3$-invariant basis in $\Sigma^3$. To further
fix notation we recall that these background
equations can be derived from the effective action \cite{2}
\begin{equation}
S_{eff}=\int d^4x
\sqrt{-g}e^{\phi}(R-\frac{1}{12}
H_{\mu\nu\rho}H^{\mu\nu\rho}+\partial_{\mu}\phi\partial^{\mu}\phi-\Lambda)
\label{b},
\end{equation}
and in the so set `sigma-' conformal frame they are
\begin{eqnarray}
R_{\mu\nu}-\frac{1}{4}H_{\mu\nu}^2-\nabla_\mu\nabla_\nu\phi&=&0,
\label{b1}\\
\nabla^2(e^\phi H_{\mu\nu\lambda})&=&0, \label{b2} \\
-R+\frac{1}{12}H^2+2\nabla^2
\phi+(\partial_\mu\phi)^2+\Lambda&=&0. \label{b3}
\end{eqnarray}
The cosmological constant $\Lambda$ (coming from a
central charge deficit in the original theory) will be
hereafter set equal to zero.
In addition to the gravitational field $g_{\mu\nu}$, expressed through
$a_i$ in (\ref{met}), these expressions also involve
the dilaton $\phi$ and,
in the contractions $H_{\mu\nu}^2=H_{\mu\kappa\lambda}
{H_{\nu}}^{\kappa\lambda}
\, , H^2=H_{\mu\nu\lambda}H^{\mu\nu\lambda}$, the totally antisymmetric
field strength $H_{\mu\nu\lambda}$. The latter, which may be equivalently
viewed here as a closed 3-form $H$, is defined in terms of the potential
$B_{\mu\nu}$ (equivalently the 2-form $B$) as
\begin{eqnarray}
H_{\mu\nu\rho}& = &\partial_\mu
B_{\nu\rho}+\partial\rho B_{\mu\nu}+\partial_{\nu} B_{\rho\mu} \label{H} \\
 (H & = & dB). \nonumber
\end{eqnarray}

Just like the metric (\ref{met}),
the dilaton and $H$ fields
must also respect the $G_3$-isometries, namely their Lie derivatives
with respect to any Killing vector
generated by the $\{\xi_i\}$ basis must vanish. This means that
the dilaton field must be a constant on $\Sigma^3$, namely it
can only be a function of the time $t$ in $M^4$.
On the other hand, the dual $H^\ast$ of $H$ must be of the form
\begin{equation}
H^\ast=H^{\ast}_0(t)dt + H^{\ast}_i(t)\sigma^i,\label{dualh}
\end{equation}
namely with components $H^{\ast}_\mu$ at most functions of $t$
in the $\{\sigma^i\}$ basis.
The (occasionally made) claim of necessarily
vanishing $H^{\ast}_i$ components in the present context
is generally false. However, due to severe restrictions, such
components can only exist in relatively few types, as we will see.
In the following, $t$ will be profitably
expressed in terms of the coordinate time $\tau$
(and with a prime for $d/d\tau$) defined by
\begin{equation}
prime=\frac{d}{d\tau}=a^{3}e^{\phi}\frac{d}{dt} \label{tau}
\end{equation}
where
\begin{equation}
a^{3}=a_1a_2a_3 \label{vol}
\end{equation}
is the expansion factor of any co-moving volume element in $\Sigma^3$.

\section{The $X(d\nearrow)$ class of Bianchi-type string backgrounds}

As implied just above, and in contrast to the mentioned
$X(d\uparrow)$, $X(d\rightarrow)$
classes, $X(d\nearrow)$ admits fewer Bianchi types whose spacetimes
satisfy the background equations (\ref{b1}--\ref{b3}).
To investigate this, one must examine
all possible isometry groups (namely each type $X$) separately
and isolate the cases in which non-vanishing
$H^{\ast}_i$ components in (\ref{dualh}) can
survive {\em in addition} to $H^{\ast}_0$. Skipping the details we
state the (easily verifiable) result that the above
requirements can be met only for certain $G_3$-class $\cal B$ types,
in fact for $X=III,V,VI_h$. The projection of
$H^\ast$ in $\Sigma^3$ is always aligned with a particular principal
direction of anisotropy. To explicitly write down these results we
recall that the just mentioned types involve a 1-parameter family of isometry
groups $G_3$ \cite{1}, which must be considered as acting on the metric
(\ref{met}). The commutation relations of the generators of such $G_3$ (given
equivalently by their dual expressions) are
\begin{equation}
d\sigma^1=0,\;\;d\sigma^2=h\sigma^1\wedge\sigma^2,\;
\;d\sigma^3=\sigma^1\wedge\sigma^3. \label{iso}
\end{equation}
The values $h=0,1$ of the real parameter $h$ give rise to
Bianchi types $III,V$ respectively, otherwise $VI_h$ is realized.
Choosing a specific realization of these $\sigma^i$, we can establish
that the possible metrics in the $X(d\rightarrow)$
class can be written as
\begin{equation}
ds^2=-dt^2+a_1^2(t)(dx^1)^2+a_2^2(t)(e^{hx^1}dx^2)^2+
a_3^2(t)(e^{x^1}dx^3)^2. \label{met6}
\end{equation}
Comparing with
(\ref{met}) one can read off (\ref{met6}) the $\sigma^i$
in terms of the $x^i$ coordinates. In fact the metric (\ref{met6})
is unique up to (the only allowed but unimportant) gauge or coordinate
transformations which preserve (\ref{iso}). We can proceed to
find the 2-form $B$ potential in (\ref{H}) in terms of the above
coordinates. The result of this calculation
may be expressed as
\begin{equation}
B=\left\{\begin{array}{ll}
\frac{1}{h+1}\eta^{\prime}(\tau)e^{(h+1)x^1}dx^2\wedge dx^3 &
|h\neq -1 \\
(A_0x^1+\zeta^{\prime}(\tau))dx^2\wedge dx^3 & |h=-1.
\end{array} \right.\label{B6}
\end{equation}
The functions $\eta(\tau)$, $\zeta(\tau)$ are to be specified and they
have been introduced through their $\tau$ derivatives for later
convenience.
It can be subsequently verifed that (as earlier claimed) the dual
$H^\ast$ in (\ref{dualh}) may be expressed as
\begin{equation}
H^\ast=a^3H^{\ast}_0(t)e^\phi d\tau- A_1e^{-\phi}\sigma^1, \label{dual1}
\end{equation}
with
\begin{equation}
a^3H^{\ast}_0=\left\{\begin{array}{ll}
-\eta^\prime & \;\;|h\neq -1 \\
-A_0 & \;\;|h=-1
\end{array} \right. \label{hast0}
\end{equation}
and with $A_0,A_1$ constants.

The background equations (\ref{b1}--\ref{b3})
can  now be given explicitly. In particular the `$ii$'
components in the set (\ref{b1}) are
\begin{eqnarray}
(\ln a_1^2e^\phi)^{\prime \prime}-2(h^2+1)(a_2a_3e^\phi)^2 & = &
 (A_1a_2a_3)^2 \nonumber \\
(\ln a_2^2e^\phi)^{\prime \prime}-2h(h+1)(a_2a_3e^\phi)^2 & = & 0
\nonumber \\
(\ln a_1^2e^\phi)^{\prime \prime}-2\;(h+1)(a_2a_3e^\phi)^2 & = & 0,
\label{b16}
\end{eqnarray}
subject to the constraint equation (the `$01$' in (\ref{b1}))
\begin{equation}
a_3^{-(h+1)}a_2^ha_1=e^{-\frac{1}{2}A_1\eta}, \label{con}
\end{equation}
plus the initial value equation (essentially the `$00$' in (\ref{b1}))
\begin{eqnarray}
(\ln a_1^2e^\phi)^{\prime}(\ln a_2^2e^\phi)^{\prime} & + &
(\ln a_2^2e^\phi)^{\prime}(\ln a_3^2e^\phi)^{\prime}  +
(\ln a_3^2e^\phi)^{\prime}(\ln a_1^2e^\phi)^{\prime} = \label{ive} \\
& = & \phi^{\prime 2}+4(h^2+h+1)(a_2a_3e^\phi)^2+
(H^\ast_0a^3e^\phi)^2+(A_1a_2a_3)^2. \nonumber
\end{eqnarray}
Coupled with (\ref{b2}),(\ref{b3}) these equations admit the following
solutions, depending on the value of the parameter $h$.
\vspace{.2cm}

{\bf For}
{\boldmath $III(d\nearrow),V(d\nearrow),VI_h(d\nearrow)$},
realized at $h=0,h=1,h\neq 0,\pm 1$, respectively, we find
\begin{eqnarray}
a_1^2e^\phi & = & Q^{\frac{2(h-1)}{h+1}}
\left(\frac{h+1}{P_1}\sinh P_1\tau\right)^{\frac{-2(h^2+1)}{(h+1)^2}}
\exp {\left(\frac{A_1}{h+1}\eta+\frac{h-1}{h+1}P_2\tau\right)} \nonumber \\
a_2^2e^\phi & = & Q^2\left(\frac{h+1}{P_1}\sinh P_1\tau\right)^
{\frac{-2h}{h+1}}\exp {(P_2\tau)} \nonumber \\
a_3^2e^\phi & = & Q^{-2}\left(\frac{h+1}{P_1}
\sinh P_1\tau\right)^{\frac{-2}{h+1}}\exp {(-P_2\tau)} \label{b166}
\end{eqnarray}
where $P_1,P_2,Q$ are constants ($Q$ could be assigned any positive value).
{}From (\ref{b3}),(\ref{b2}) for the dilaton and $H$ field we
obtain the coupled system
\begin{eqnarray}
\phi^{\prime \prime} & = & \left(\frac{A_1P_1}{h+1}\right)^2
(e^\phi\sinh P_1\tau)^{-2}-e^{2\phi}\eta^{\prime 2} \label{phi} \\
\eta^{\prime \prime} & = & \frac{A_1P_1^2}{h+1}
(e^\phi\sinh P_1\tau)^{-2} \label{eta}
\end{eqnarray}
and hence the functions $\phi(\tau),\eta(\tau)$, although apparently not in
closed form in the general case. They are subject to the initial value
equation
\begin{equation}
\phi^{\prime 2}+e^{2\phi}\eta^{\prime 2}+\frac{2A_1P_1}{h+1}(\coth P_1\tau)
\eta^{\prime}+\left(\frac{A_1P_1}{h+1}\right)^2(e^\phi\sinh P_1\tau)^{-2} =
4\frac{h^2+h+1}{(h+1)^2}P_1^2-P_2^2, \label{ive6}
\end{equation}
as required by (\ref{ive}). The
constraint (\ref{con}) has already been satisfied in view of (\ref{b166}).
At the $A=0$ limit, which according to (\ref{dual1}) corresponds to
a hypersurface-orthogonal $H^\ast$, the above set can be easily integrated
to reproduce the already known $III(d\uparrow)$,$V(d\uparrow)$ and
$VI_h(d\uparrow)\;$ cases \cite{6}.
A multitude of other special cases is possible. We explicitly mention
the $V(d\nearrow)$, with asymptotic  $SO(3)$ isotropy,
realized as that spacetime expands towards an open $(k=-1)$ FRW model
according to (\ref{b166}--\ref{ive6}) at $h=1$.
\vspace{.2cm}

{\bf For}
{\boldmath $VI_{-1}(d\nearrow)$},
realized at $h=-1$, we find
\begin{eqnarray}
a_1^2e^\phi & = & Q_1^2e^{P_1\tau}\exp (A_1\zeta +Q_1^2Q_2^2e^{2P_2\tau})
 \nonumber \\
a_2^2e^\phi & = & Q_2^2|P_2|\exp {(P_2+A_0A_1/2)\tau} \nonumber \\
a_3^2e^\phi & = & Q_3^2|P_2|\exp {(P_2-A_0A_1/2)\tau}  \label{b160}
\end{eqnarray}
where $Q_i,P_1,P_2$ are constants. From (\ref{b3}),(\ref{b2})
for the dilaton and $H$ field we obtain the coupled system
\begin{eqnarray}
\phi^{\prime \prime} & = & -A_0^2e^{2\phi}+(Q_0/A_0)^2e^{2P_2\tau-2\phi}
\label{phi1} \\
A_1\zeta^{\prime \prime} & = & (Q_0/A_0)^2e^{2P_2\tau-2\phi} \label{zeta}
\end{eqnarray}
subject to the initial value equation
\begin{equation}
\phi^{\prime 2}-2A_1P_2\zeta^{\prime}+A_0^2e^{2\phi}+
(Q_0/A_0)^2e^{2P_2\tau-2\phi}=P_2^2+2P_1P_2-\frac{1}{4}(A_1A_2)^2 \label{ive1}
\end{equation}
as follows from (\ref{ive}), with
$Q_0=|A_0A_1Q_2Q_3P_2|$.
As in the previous case, the general solution for the functions
$\phi(\tau),\zeta(\tau)$ does not seem attainable in closed form.
Here, however, we observe that the dilaton field may be expressed as
\begin{equation}
e^{2\phi}=A_0^{-2}Q_0e^{\psi+P_2\tau}, \label{f}
\end{equation}
where $\psi(\tau)$ is a solution to
\begin{equation}
\psi^{\prime \prime}+4Q_0e^{P_2\tau}\sinh \psi. \label{psi}
\end{equation}
For positive $P_2$, one can
interpret the `sinh' term as a confining potential
(in relation to the $\psi=0$ convergence limit)
to realize that $\psi$ must
oscillate around zero with exponentially {\em increasing} frequencies and
{\em decreasing} amplitudes.
Setting $\psi=0$ in
(\ref{f}) etc., we find
\begin{eqnarray}
e^{2\phi} &=& A_0^{-2}Q_0e^{P_2\tau}, \label{f0} \\
A_1\zeta &=& (Q_0/P_2^2)e^{P_2\tau}. \label{z0}
\end{eqnarray}
The rest of the solution is given by (\ref{b160}),
together with the restriction
\begin{equation}
3P_2^2+8P_1P_2=(A_1A_2)^2 \label{ic}
\end{equation}
coming from (\ref{ive1}).
The result expressed by (\ref{b160}) together with
(\ref{f0}--\ref{ic}) does {\em not} give the most general
$VI_{-1}(d\nearrow)$ possible, but rather the asymptotic limit of the
general solution. The same result is also by itself a solution (in closed
form) and, as such, it essentially reproduces as special cases
each one of the $VI_{-1}(d\uparrow)$ and $VI_{-1}(d\rightarrow)$ solutions
found in \cite{6} and \cite{7} respectively.

\section{Conclusions}

\newcommand{\e}{$\boldmath\exists$}
\newcommand{\n}{$\not\!\exists$}
\newcommand{\ep}{$\exists^\ast$}
\newcommand{\ri}{$\;\Rightarrow$}

All possible BTSBs may be classified, so that each one is represented
as $X^n(d,a)$. The parameter $X$ specifies the isometry group $G_3$
and takes the values $I,II,\ldots,IX$, roughly one for
each of the Bianchi types plus one for the Kantowski-Sachs
class of metrics \cite{1}. The rest of the parameters may
be omitted or take values as follows.
The index $n$ specifies the isotropy group, which
is $SO(3)$ if $n=3$, $SO(2)$ if $n=2$, or the null group (case of
complete anisotropy) if $n$ is omitted altogether. The argument $d$ is
omitted only when the metric (\ref{met}) is
{\em non-}diagonal in the $\{\sigma^i\}$ basis.
The last argument, zero in the trivial
case of an identically vanishing $H$ field,
specifies the orientation of $H^\ast$ relative to
the hypersurfaces of homogeneity $\Sigma^3$.
There are no classification parameters for the
dilaton field
and the cosmological constant $\Lambda$.
With types $VI_{-1}$ and $VII_0$ counted separately,  it turns out that
there are in all $24^2=576$ cases which this classification sees as distinct
BTSBs. Many of them are obviously special cases, descending from more
general (less-symmetric) ones, as, for example, the `\ri' arrows in the Table
indicate. Others (such as the `\n ` cases
in the Table) cannot
be realized in the sense that their metrics are singular
everywhere. This by no means implies that the respective geometries
do not exist.  It does mean, however, that such
manifolds could {\em in principle} be realized
only in the presence of appropriate sources.

Before turning to specific Bianchi types, we will briefly review
certain aspects common to all cases.
Let us begin by taking as an example the Bianchi-type $V$ case,
for which we copy from the Table the sequence
\begin{equation}
\cdots\;\; V(d\nearrow)\;\;\Rightarrow \;\;V(d\uparrow)\;\;
\Rightarrow \;\; V^2(d\uparrow) \;\;\Rightarrow \;\; V^3(d\uparrow).
\end{equation}
Any spatial component of $H^\ast$ breaks $SO(3)$ isotropy,
so that FRW behavior (if at all possible) can exist only
for vanishing $H{^\ast}_i$. Clearly, one expects special interest in the
cases of {\em asymptotic} attainment of this value. In the $V(d\nearrow)$
case, $H^\ast$ is tangent
to a congruence which could sustain general kinematics, namely
expansion, shear (anisotropy) and, {\em in principle}, vorticity as well.
In the $V(d\uparrow)$ case, $SO(3)$ isotropy may still be broken by other
agents but vorticity must vanish identically, as it
cannot be sustained for kinematical reasons.
In the terminal $V^3(d\uparrow)$ case, namely the open FRW model,
only isotropic expansion has survived.
One can further establish
that in all cases there is an initial singularity and no inflation.

\vspace{.2cm}
{\bf Type I\@.\/}
The isotropy limit contains
all possible flat ($k=0$) FRW cases. Until recently, only the $I^3(d0)$
with its $I^3(d\uparrow)$ and the Kasner-like $I(d0)$
generalization had been given\cite{3},\cite{4},\cite{5}.
They are all reproduced as special cases of
$I(d\uparrow)$ \cite{6}.
The $I(d\rightarrow)$ is also known  \cite{7}, but, according to our result
covering all $G_3$-class $\cal A$ types, there exists no $I(d\nearrow)$.

{\bf Type II\@.\/}
The fully anisotropic $II(d\uparrow)$ given in \cite{6}
generalizes the $II(d0)$ found in \cite{5}.
For the $II(d\rightarrow)$ and $II(d\nearrow)$
cases the same hold as for type I.

{\bf Type III\@.\/}
$III(d\nearrow)$ exists, as we saw, and it reduces
to the $III(d\uparrow)$ found in \cite{6}.
However, there also exists a general $III(d\rightarrow)$ \cite{7},
which cannot be reached as a limit of
the mentioned $III(d\nearrow)$.

{\bf Type IV\@.\/}
All diagonal metrics are singular everywhere \cite{6},\cite{7}.

{\bf Type V\@.\/}
At isotropy one obtains all  open ($k=-1$) FRW cases, such as
the $V^3(d0)$ and $V^3(d\uparrow)$
found in \cite{3}, \cite{4}, the first one
generalized by the $V(d0)$ in \cite{5}. They all are
special cases of $V(d\uparrow)$, given in \cite{6},
with the latter further generalized by the  $V(d\nearrow)$ found here.
There exists no $V(d\rightarrow)$ case \cite{7}.

{\bf Type $VI_h$\@.\/}
The results of the previous case are generally valid here as well,
except of course for the isotropy limit.

{\bf Type $VI_{-1}$\@.\/}
Generally, the $h=-1$ case is {\em not} obtained
at the $h=-1$ limit from solutions of the
previous type. We have seen that $VI_{-1}(d\nearrow)$ generalizes
the $VI_{-1}(d\uparrow)$ in \cite{6} as well as the
$VI_{-1}(d\rightarrow)$ \cite{7}. This is the only case in which
{\em both} such limits can be reached from a given $X(d\nearrow)$.

{\bf Type $VII_h$\@.\/}
In this case, which also involves a 1-parameter group $G_3$
(here with $h^2\leq 4$),
all metrics are singular  everywhere, unless $h=0$.
The latter case (which exceptionally involves
spacetimes of  $G_3$-class $\cal A$) is outlined next.

{\bf Type $VII_0$\@.\/}
There exists the $VII_0(d\uparrow)$ and its isotropy
limits $VII_0^2(d\uparrow)$ and $VII_0^3(d\uparrow)$
are identical to those in the Type-I case \cite{6}.
The $VII_0(d\rightarrow)$ has been given recently \cite{7}
but, as we have seen, there exists no $VII_0(d\nearrow)$.

{\bf Type VIII\@.\/}
We have seen that there exists no $VIII(d\nearrow)$.
Neither is there a $VIII(d\rightarrow)$  \cite{7},
but the $VIII^2(d\uparrow)$ has been explicitly found in \cite{6}.

{\bf Type IX\@.\/}
Also explicitly found is the $IX^2(d\uparrow)$ case
(generalizing the well-known Taub metric to which it reduces) \cite{6}.
The complete isotropy limit therein, namely $IX^3(d\uparrow)$,
reproduces the closed $(k=1)$ FRW cases found in \cite{3},\cite{4}.
In view of our present findings (cf. also \cite{7}), the most general
possible diagonal Bianchi-type $IX$ case is $IX(d\uparrow)$
(but elusive just like its Mixmaster counterpart!).

To conclude with some generally applicable remarks, we note that
the {\em energy of anisotropy} \cite{1},\cite{6}
may be quite significant, compared with that of
any other field present in the effective action (\ref{b}), during some
time near the Planck or string scale. It follows that the study of
anisotropy in such strong-field regimes would
require solutions which are exact to all orders
in the $\alpha^\prime$ organization of the string action.
It is apparently not known whether some (all?) of the more general
$X(d)$ backgrounds discussed here
(that is, with no more than three Killing isometries)
are exact solutions beyond leading order
in $\alpha^\prime$. Relevant in that context,
although applied to a different class of
solutions, is ref. \cite{8}.
We also note that, under abelian target space duality, the {\em known}
solutions in $X(d\uparrow)$ generate metrics in the same class \cite{6}.
Obviously, however, this cannot be the case in general. For example,
inspection of (\ref{B6}) etc., immediately shows that the duals
of $VI_h(d\nearrow)$ will involve
metrics with non-vanishing `$0i$' components in the
invariant basis. With the latter type of metric, even if homogeneity
were not really lost, there would be  no universal time to define uniquely
the (presumably observable) 3D sections of homogeneity. Such spacetimes
appear to be of significance very close to the initial singularity \cite{9}.

\vspace{1cm}
I would like to thank  the Theory Division at CERN for hospitality.

\newpage
\leftmargin=0cm
\begin{center}
{\bf The diagonal Bianchi-type string backgrounds }
\end{center}

\begin{center}
\begin{tabular}{|c|c|c|c|c|c|c|c|} \hline
$X$ & $d\sigma^i=$ & $G_3$ & $X(d\rightarrow)$ & X($d\nearrow$) &
$X(d\uparrow$) & $X^2(d\uparrow)$ & $X^3(d\uparrow)$ \\
 &$\frac{1}{2}C^i_{jk}\sigma^j\wedge \sigma^k$ & class & & & & &(FRW) \\
\hline \hline

 & & & & & & & \\
$I$& $d\sigma^i=0$&$\cal A$&\e &\n &\e \ri &\e \ri &\e \\
 & & & & & & & \\ \hline
 &$d\sigma^1=\sigma^2\wedge \sigma^3$& & & & & & \\
$II$&$d\sigma^2=0$&$\cal A$ &\e &\n &\e &\n &\n \\
 &$d\sigma^3=0$& & & & & &\\ \hline
 &$d\sigma^1=0$& & & & & & \\
$III$&$d\sigma^2=0$&$\cal B$&\e &\e \ri &\e &\n &\n \\
 &$d\sigma^3=\sigma^1\wedge \sigma^3$& & & & &  & \\ \hline
&$d\sigma^1=\sigma^1\wedge \sigma^3+$& & & & & &\\
$IV$&$\sigma^2\wedge \sigma^3,\,d\sigma^2=$&$\cal B$&\n&\n&\n&\n&\n\\
&$\sigma^2\wedge \sigma^3,\,d\sigma^3=0$& & & & & & \\ \hline
&$d\sigma^1=0$& &  & & & &\\
$V$&$\sigma^2=\sigma^1\wedge \sigma^2$&$\cal B$&
\n &\e \ri &\e \ri &\e \ri &\e \\
&$d\sigma^3=\sigma^2\wedge \sigma^3$&  & & & & & \\ \hline
&$d\sigma^1=0$& & & & & & \\
$VI_h$&$\sigma^2=
h\sigma^1\wedge \sigma^2$&$\cal B$&\n &\e \ri &\e &\n &\n \\
&$d\sigma^3=\sigma^1\wedge \sigma^3$& & & & & &\\ \cline{3-8}
 & & & & & & & \\
$VI_{-1}$&$(h=-1)$&$\cal B$&\e &$\Leftarrow$\e \ri &\e &\n& \n \\
 & & & & & & & \\ \hline
&$d\sigma^1=-\sigma^2\wedge \sigma^3$& & & & & &\\
$VII_h$&$d\sigma^2=\sigma^1\wedge \sigma^3+$&$\cal B$&\n &\n &\n &\n &\n \\
&$h\sigma^2\wedge \sigma^3,\,d\sigma^3=0$& & & & & & \\ \cline{3-8}
 & & & & & & & \\
$VII_0$&$(h=0)$&$\cal A$&\ep &\n &\ep \ri &\e \ri &\e \\
 & & & & & & & \\ \hline
&$d\sigma^1=\sigma^2\wedge \sigma^3$& & & & & & \\
$VIII$&$d\sigma^2=-\sigma^3\wedge \sigma^1$
&$\cal A$&\n &\n &\ep \ri &\e &\n \\
&$d\sigma^3=\sigma^1\wedge \sigma^2$& & & & & & \\ \hline
&$d\sigma^1=\sigma^2\wedge \sigma^3$& & & & & & \\
$IX$&$d\sigma^2=\sigma^3\wedge \sigma^1$&$\cal A$&\n &\n &
\ep \ri &\e \ri &\e \\
&$d\sigma^3=\sigma^1\wedge \sigma^2$& & & & & &\\ \hline
\end{tabular}
\end{center}

\noindent
\e  :$\;$ solution is known (in some cases not in entirely closed form,
see also \cite{6},\cite{7}).

\noindent
\ep :$\;$ solution exists but it is not known.

\noindent
\n  :$\;$ not-realizable as a `vacuum' background.

\noindent
\ri :$\;$ towards specialization (higher symmetry).

\end{document}